# Developing a Visual Interactive Search History Exploration System

**Wilko van Hoek**
GESIS - Leibniz Institute for the Social Sciences
Cologne, Germany
wilko.vanhoek@gesis.org

## MOTIVATION

Visualizing search histories and supplying the user with interaction methods such as zooming, brushing and linking, annotation and tagging, has attracted increasing attention in scientific literature (Liu, Hong, & Pedersen, 2010), (Pedersen, Gyllstrom, Gu, & Hong, 2010), (Cernea, Truderung, Kerren, & Ebert, 2013), and (Yu & Ingalls, 2011). The overall assumption is that the user's information seeking process often is divided into separate sessions, and that search histories can help finding the point of discontinuation and ease resuming the interrupted process. The several search sessions that belong together are defined as trails.

Resuming discontinued trails is not the only potential that lies in search histories. Over various searches users view, rate, and discard various different documents and conduct multiple searches. In digital libraries the search trail represents a subset of the library and thus can be viewed as an individual library. In contrast to the actual digital library, the search history only consists of documents which have been viewed by the user and thus is qualified for re-finding tasks. (Morris, Ringel Morris, & Venolia, 2008) find that their system, which provided a hierarchical search history, is suitable for re-finding information and task reacquisition.

Instead of strictly viewing search histories as a means to the end of re-finding documents, we propose to interpret the search history as an individual library and thus as the user's knowledge base. We will develop a visual and interactive search history exploration system that transfers state-of-the-art functionalities for digital libraries into a search history. In addition we will implement further methods for organizing and individualizing the search history.

## METHODOLOGY

Along a series of user studies we will identify suitable concepts to support search history exploration and develop a prototypical system for search history exploration. The studies will be divided into three phases. In the first phase we will assess usability of different visualization and interaction techniques in small experiments in a laboratory environment, to identify suitable combinations of visualizations and interactions. In the second phase we will implement the most usable combinations and evaluate their performance regarding different tasks in a new experiment again in a laboratory environment. Based on the results we will refine the prototype and distribute it to DLs such as Sowiport (http://www.gesis.org/sowiport) related-work.net (http://related-work.net), SSOAR (http://www.ssoar.info) and ezDL (http://www.ezdl.dw). We will evaluate the different users' behaviours in a longitudinal study over a period of several weeks up to a few months.

## VISUALIZATION AND INTERACTION TECHNIQUES SUPPORTING SEARCH HISTORY EXPLORATION

Search histories pose different visualization possibilities. They are structured according to the timestamps of the actions leading to a certain object and they possess internal relations such as

search <-> search result, author <-> paper. Based on the relational interpretation they could be visualized as a network-graph or a hierarchical tree (cf. Figure 1). For our prototype we will evaluate at least three types of visualizations (hierarchical tree-list, hierarchical tree and network-graph) regarding their suitability to support search history exploration. We also consider other types of visualizations we were able to classify in (van Hoek & Mayr, 2013).

Basic interaction techniques for search history exploration in a visual and interactive system are searching, filtering, sorting, brushing and linking, and zooming. To individualize the search history, users should be able to annotate, tag, and cluster the various objects/events. An important feature could be the possibility to cluster documents and tag them, to be able to filter the search history in regard to a certain topic in the future. To make histories reusable they should also be savable as well as loadable. The saving and loading process should support sharing of search histories and also parts of it. Another field of functionalities that should be considered is collaboration. Different users should be able to merge/join their histories to work on it together.

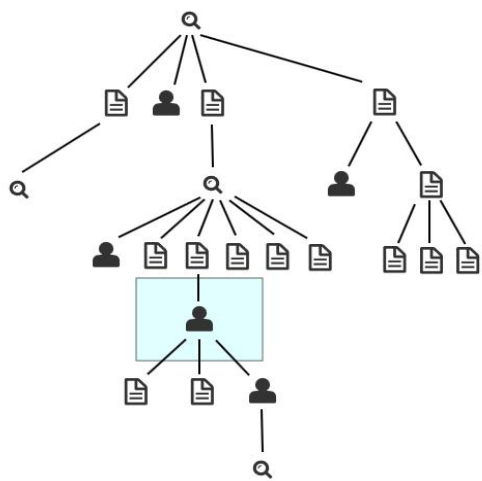

*Figure 1-A search history visualized as a hierarchical tree.*